\documentclass[10pt,a4paper,onecolumn]{article}
\usepackage{marginnote}
\usepackage{graphicx}
\usepackage[rgb]{xcolor}
\usepackage{authblk,etoolbox}
\usepackage{titlesec}
\usepackage{calc}
\usepackage{tikz}
\usepackage[pdfa]{hyperref}
\usepackage{hyperxmp}
\hypersetup{%
    unicode=true,
    pdfapart=3,
    pdfaconformance=B,
    pdftitle={rustworkx: A High-Performance Graph Library for Python},
    pdfauthor={Matthew Treinish, Ivan Carvalho, Georgios
Tsilimigkounakis, Nahum Sá},
    pdfpublication={Journal of Open Source Software},
    pdfpublisher={Open Journals},
    pdfissn={2475-9066},
    pdfpubtype={journal},
    pdfvolumenum={},
    pdfissuenum={},
    pdfdoi={10.21105/joss.03968},
    pdfcopyright={Copyright (c) 1970, Matthew Treinish, Ivan Carvalho,
Georgios Tsilimigkounakis, Nahum Sá},
    pdflicenseurl={http://creativecommons.org/licenses/by/4.0/},
    colorlinks=true,
    linkcolor=[rgb]{0.0, 0.5, 1.0},
    citecolor=Blue,
    urlcolor=[rgb]{0.0, 0.5, 1.0},
    breaklinks=true
}
\immediate\pdfobj stream attr{/N 3} file{sRGB.icc}
\pdfcatalog{%
  /OutputIntents [
    <<
      /Type /OutputIntent
      /S /GTS_PDFA1
      /DestOutputProfile \the\pdflastobj\space 0 R
      /OutputConditionIdentifier (sRGB)
      /Info (sRGB)
    >>
  ]
}
\usepackage{caption}
\usepackage{orcidlink}
\usepackage{tcolorbox}
\usepackage{amssymb,amsmath}
\usepackage{ifxetex,ifluatex}
\usepackage{seqsplit}
\usepackage{xstring}

\usepackage{float}
\let\origfigure\figure
\let\endorigfigure\endfigure
\renewenvironment{figure}[1][2] {
    \expandafter\origfigure\expandafter[H]
} {
    \endorigfigure
}

\usepackage{fixltx2e} % provides \textsubscript

\newlength{\cslhangindent}
\newlength{\csllabelwidth}
\newenvironment{CSLReferences}[2] % #1 hanging-ident, #2 entry spacing
 {% don't indent paragraphs
  \setlength{\parindent}{0pt}
  \ifodd #1 \everypar{\setlength{\hangindent}{\cslhangindent}}\ignorespaces\fi
  \ifnum #2 > 0
  \setlength{\parskip}{#2\baselineskip}
  \fi
 }%
 {}
\usepackage{calc}

\usepackage[top=3.5cm, bottom=3cm, right=1.5cm, left=1.31cm,
            headheight=2.2cm, reversemp, includemp, marginparwidth=4.1cm]{geometry}

\titleformat{\section}
  {\normalfont\sffamily\Large\bfseries}
  {}{0pt}{}
\titleformat{\subsection}
  {\normalfont\sffamily\large\bfseries}
  {}{0pt}{}
\titleformat{\subsubsection}
  {\normalfont\sffamily\bfseries}
  {}{0pt}{}
\titleformat*{\paragraph}
  {\sffamily\normalsize}

\usepackage{fancyhdr}
\makeatletter
\let\ps@plain\ps@fancy
\fancyheadoffset[L]{4.5cm}
\fancyfootoffset[L]{4.5cm}

\definecolor{linky}{rgb}{0.0, 0.5, 1.0}

\newtcolorbox{repobox}
   {colback=red, colframe=red!75!black,
     boxrule=0.5pt, arc=2pt, left=6pt, right=6pt, top=3pt, bottom=3pt}

\newcommand{\ExternalLink}{%
   \tikz[x=1.2ex, y=1.2ex, baseline=-0.05ex]{%
       \begin{scope}[x=1ex, y=1ex]
           \clip (-0.1,-0.1)
               --++ (-0, 1.2)
               --++ (0.6, 0)
               --++ (0, -0.6)
               --++ (0.6, 0)
               --++ (0, -1);
           \path[draw,
               line width = 0.5,
               rounded corners=0.5]
               (0,0) rectangle (1,1);
       \end{scope}
       \path[draw, line width = 0.5] (0.5, 0.5)
           -- (1, 1);
       \path[draw, line width = 0.5] (0.6, 1)
           -- (1, 1) -- (1, 0.6);
       }
   }

\patchcmd{\@maketitle}{center}{flushleft}{}{}
\patchcmd{\@maketitle}{center}{flushleft}{}{}
\patchcmd{\@maketitle}{\LARGE}{\LARGE\sffamily}{}{}
\def\maketitle{{%
  
  \AB@maketitle}}
\renewcommand\AB@affilsepx{ \protect\Affilfont}
\renewcommand\AB@affilnote[1]{{\bfseries #1}\hspace{3pt}}
\renewcommand{\affil}[2][]%
   {\newaffiltrue\let\AB@blk@and\AB@pand
      \if\relax#1\relax\def\AB@note{\AB@thenote}\else\def\AB@note{#1}%
        \setcounter{Maxaffil}{0}\fi
        \begingroup
        \let\href=\href@Orig
        \let\protect\@unexpandable@protect
        \def\thanks{\protect\thanks}\def\footnote{\protect\footnote}%
        \@temptokena=\expandafter{\AB@authors}%
        {\def\\{\protect\\\protect\Affilfont}\xdef\AB@temp{#2}}%
         \xdef\AB@authors{\the\@temptokena\AB@las\AB@au@str
         \protect\\[\affilsep]\protect\Affilfont\AB@temp}%
         \gdef\AB@las{}\gdef\AB@au@str{}%
        {\def\\{, \ignorespaces}\xdef\AB@temp{#2}}%
        \@temptokena=\expandafter{\AB@affillist}%
        \xdef\AB@affillist{\the\@temptokena \AB@affilsep
          \AB@affilnote{\AB@note}\protect\Affilfont\AB@temp}%
      \endgroup
       \let\AB@affilsep\AB@affilsepx
}
\makeatother

\renewcommand\Affilfont{\sffamily\small\mdseries}
\ifnum 0\ifxetex 1\fi\ifluatex 1\fi=0 % if pdftex
  \usepackage[T1]{fontenc}
  \usepackage[utf8]{inputenc}

\else % if luatex or xelatex
  \ifxetex
    \usepackage{mathspec}
    \usepackage{fontspec}

  \else
    \usepackage{fontspec}
  \fi
  \defaultfontfeatures{Ligatures=TeX,Scale=MatchLowercase}

\fi
\IfFileExists{upquote.sty}{\usepackage{upquote}}{}
\IfFileExists{microtype.sty}{%
\usepackage{microtype}
\UseMicrotypeSet[protrusion]{basicmath} % disable protrusion for tt fonts
}{}

\PassOptionsToPackage{usenames,dvipsnames}{color} % color is loaded by hyperref
\ifLuaTeX
\usepackage[bidi=basic]{babel}
\else
\usepackage[bidi=default]{babel}
\fi
\babelprovide[main,import]{american}
\def\languageshorthands#1{}

\usepackage{graphicx,grffile}
\makeatletter
\def\maxwidth{\ifdim\Gin@nat@width>\linewidth\linewidth\else\Gin@nat@width\fi}
\def\maxheight{\ifdim\Gin@nat@height>\textheight\textheight\else\Gin@nat@height\fi}
\makeatother
\setkeys{Gin}{width=\maxwidth,height=\maxheight,keepaspectratio}
\IfFileExists{parskip.sty}{%
\usepackage{parskip}
}{% else
\setlength{\parindent}{0pt}
\setlength{\parskip}{6pt plus 2pt minus 1pt}
}
\ifx\paragraph\undefined\else
\let\oldparagraph\paragraph
\renewcommand{\paragraph}[1]{\oldparagraph{#1}\mbox{}}
\fi
\ifx\subparagraph\undefined\else
\let\oldsubparagraph\subparagraph
\renewcommand{\subparagraph}[1]{\oldsubparagraph{#1}\mbox{}}
\fi
\usepackage{multicol}
\ifLuaTeX
  \usepackage{selnolig}  % disable illegal ligatures
\fi

\title{rustworkx: A High-Performance Graph Library for Python}

\author[1%
]{Matthew Treinish%
  \,\orcidlink{0000-0001-9713-2875}\,%
}
\author[2%
]{Ivan Carvalho%
  \,\orcidlink{0000-0002-8257-2103}\,%
}
\author[3%
]{\newline Georgios Tsilimigkounakis%
  \,\orcidlink{0000-0001-6174-0801}\,%
}
\author[4%
]{Nahum Sá%
  \,\orcidlink{0000-0002-3234-8154}\,%
}

\affil[1]{IBM Quantum, IBM T.J. Watson Research Center, Yorktown
Heights, USA \newline}
\affil[2]{University of British Columbia, Kelowna, Canada \newline}
\affil[3]{National Technical University of Athens, Athens, Greece
\newline}
\affil[4]{Centro Brasileiro de Pesquisas Físicas, Rio de Janeiro,
Brazil}
\date{\vspace{-2.5ex}}

\begin{document}
\maketitle

\marginpar{

  \begin{flushleft}
  %\hrule
  \sffamily\small

  {\bfseries DOI:} \href{https://doi.org/10.21105/joss.03968}{\color{linky}{10.21105/joss.03968}}

  \vspace{2mm}

  {\bfseries Software}
  \begin{itemize}
    \setlength\itemsep{0em}
    \item \href{https://github.com/openjournals/joss-reviews/issues/3968}{\color{linky}{Review}} \ExternalLink
    \item \href{https://github.com/Qiskit/rustworkx}{\color{linky}{Repository}} \ExternalLink
    \item \href{https://doi.org/10.5281/zenodo.5879860}{\color{linky}{Archive}} \ExternalLink
  \end{itemize}

  \vspace{2mm}

  \par\noindent\hrulefill\par

  \vspace{2mm}

  {\bfseries Editor:} \href{https://vknight.org/}{Vincent Knight} \ExternalLink \orcidlink{0000-0002-4245-0638}
   \\
  \vspace{1mm}
    {\bfseries Reviewers:}
  \begin{itemize}
  \setlength\itemsep{0em}
    \item \href{https://github.com/szhorvat}{@szhorvat}
    \item \href{https://github.com/inakleinbottle}{@inakleinbottle}
    \end{itemize}
    \vspace{2mm}

  {\bfseries Submitted:} 28 October 2021\\
  {\bfseries Published:} 01 November 2022

  \vspace{2mm}
  {\bfseries License}\\
  Authors of papers retain copyright and release the work under a Creative Commons Attribution 4.0 International License (\href{https://creativecommons.org/licenses/by/4.0/}{\color{linky}{CC BY 4.0}}).

  \end{flushleft}
}

\begin{quote}
\begin{quote}
In \emph{\href{https://github.com/Qiskit/rustworkx}{rustworkx}}, we
provide a high-performance, flexible graph library for Python.
\emph{rustworkx} is inspired by \emph{NetworkX} but addresses many
performance concerns of the latter. \emph{rustworkx} is written in Rust
and is particularly suited for performance-sensitive applications that
use graph representations.
\end{quote}
\end{quote}

\hypertarget{statement-of-need}{%
\section{Statement of need}\label{statement-of-need}}

\emph{rustworkx} is a general-purpose graph theory library focused on
performance. It wraps low-level Rust code
(\protect\hyperlink{ref-Matsakis2014}{Matsakis \& Klock, 2014}) with a
flexible Python API, providing fast implementations for graph data
structures and popular graph algorithms.

\emph{rustworkx} is inspired by the \emph{NetworkX} library
(\protect\hyperlink{ref-SciPyProceedings_11}{Hagberg et al., 2008}), but
meets the needs of users that also need performance. Even though
\emph{NetworkX} is the de-facto standard graph and network analysis
library for Python, it has performance concerns. \emph{NetworkX}
prefers pure Python implementations, which leads to bottlenecks in
computationally intensive applications that use graph algorithms.

\emph{rustworkx} addresses those performance concerns by switching to a
Rust implementation. It has support for shortest paths, isomorphism,
matching, multithreading via rayon
(\protect\hyperlink{ref-Stone2021}{Stone \& Matsakis, 2021}), and much
more.

\hypertarget{related-work}{%
\section{Related work}\label{related-work}}

The graph and network analysis ecosystem for Python is rich, with many
libraries available. \emph{igraph}
(\protect\hyperlink{ref-Csardi2006}{Csardi \& Nepusz, 2006}),
\emph{graph-tool} (\protect\hyperlink{ref-Peixoto2014}{Peixoto, 2014}),
\emph{SNAP} (\protect\hyperlink{ref-Leskovec2016}{Leskovec \& Sosič,
2016}), and \emph{Networkit} (\protect\hyperlink{ref-Staudt2016}{Staudt
et al., 2016}) are Python libraries written in C or C++ that can replace
\emph{NetworkX} with better performance. We also highlight
\emph{SageMath}’s graph theory module
(\protect\hyperlink{ref-Sagemath2020}{The Sage Developers, 2020}), which
has a stronger focus in mathematics than \emph{NetworkX}.

However, the authors found that no library matched the flexibility that
\emph{NetworkX} provided for interacting with graphs. \emph{igraph} is
efficient for static large graphs, but does not handle graph updates as
efficiently. \emph{SNAP} and \emph{Networkit} do not support associated
edge data with arbitrary Python types. \emph{graph-tool} supports
associated edge data at the cost of maintaing the data in a separate
data structure. Thus, the main contribution of \emph{rustworkx} is
keeping the ease of use of \emph{NetworkX} without sacrificing
performance.

We note that existing code using \emph{NetworkX} needs to be modified to
use \emph{rustworkx}. \emph{rustworkx} is not a drop-in replacement for
\emph{NetworkX}, which may be a possible limitation for some users. The
authors provide a \emph{NetworkX} to \emph{rustworkx} conversion guide
in the documentation to aid in those situations.

\hypertarget{graph-data-structures}{%
\section{Graph data structures}\label{graph-data-structures}}

\emph{rustworkx} provides two core data structures: \texttt{PyGraph} and
\texttt{PyDiGraph}. They correspond to undirected and directed graphs,
respectively. Graphs describe a set of nodes and the edges connecting
pairs of those nodes. Internally, \emph{rustworkx} leverages the
\emph{petgraph} library (\protect\hyperlink{ref-bluss2021}{bluss et al.,
2021}) to store the graphs using an adjacency list model and the
\emph{PyO3} library (\protect\hyperlink{ref-Hewitt2021}{Hewitt et al.,
2021}) for the Python bindings.

Nodes and edges of the graph may also be associated with data payloads.
Payloads can contain arbitrary data, such as node labels or edge
lengths. Common uses of data payloads include representing weighted
graphs. Any Python object can be a data payload, which makes the library
flexible because no assumptions are made about the payload types.

\emph{rustworkx} operates on payloads with callbacks. Callbacks are
functions that take payloads and return statically typed data. They
resemble the named attributes in \emph{NetworkX}. Callbacks are
beneficial because they bridge the arbitrary stored data with the static
types \emph{rustworkx} expects.

A defining characteristic of \emph{rustworkx} graphs is that each node
maps to a non-negative integer node index, and similarly, each edge maps
to an edge index. Those indices uniquely determine nodes and edges in
the graph. Indices are stable, hence the index for a node \(v\) does not
change even if another node \(u\) is removed. Moreover, indices separate
the data representing the graph’s structure, which is stored in Rust,
from the payloads associated with nodes and edges, which are stored in
Python.

\begin{figure}
\centering
\includegraphics[width=1\textwidth,height=\textheight]{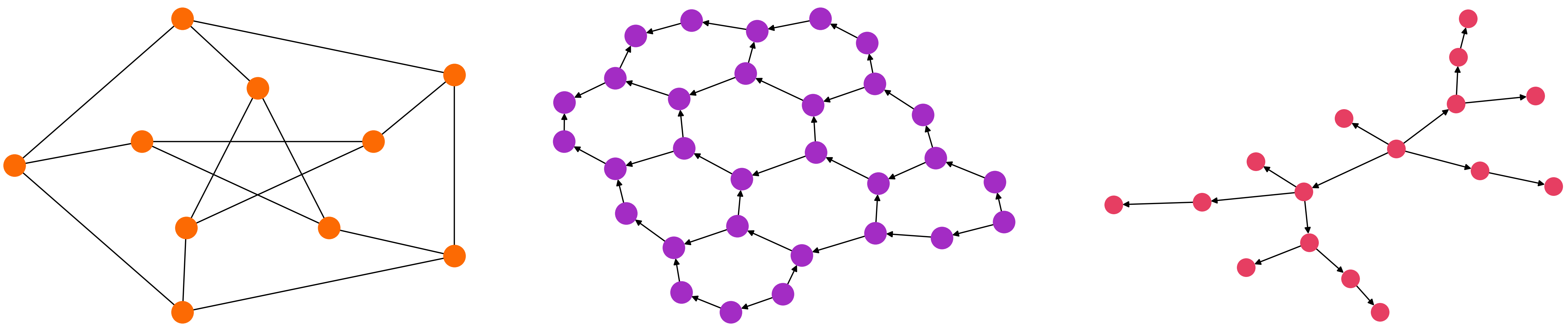}
\caption{A Petersen graph, a hexagonal lattice graph, and a binomial
tree graph created with \textbf{\texttt{rustworkx.generators}} and
visualized with the \textbf{\texttt{rustworkx.visualization}}
module.\label{fig:graphexample}}
\end{figure}

\hypertarget{use-cases}{%
\section{Use Cases}\label{use-cases}}

\emph{rustworkx} is suitable for modeling graphs ranging from a few
nodes scaling up to 4 billion. The library is particularly suited for
applications that have core routines executing graph algorithms. In
those applications, the performance of \emph{rustworkx} considerably
reduces computation time. Examples of applications using
\emph{rustworkx} include the Qiskit compiler
(\protect\hyperlink{ref-Qiskit2021}{Treinish et al., 2021}), PennyLane
(\protect\hyperlink{ref-Bergholm2020}{Bergholm et al., 2020}), atompack
(\protect\hyperlink{ref-Ullberg2021}{Ullberg, 2021}), and qtcodes
(\protect\hyperlink{ref-Jha2021}{Jha et al., 2021}).

For common use cases, \emph{rustworkx} can provide speedups ranging from
3x to 100x compared to the same code using \emph{NetworkX} while
staying competitive with other compiled libraries like \emph{igraph} and
\emph{graph-tool}. The gains in performance are application-specific,
but as a general rule, the more work that is offloaded to
\emph{rustworkx} and Rust, the larger are the gains.

We illustrate use cases with examples from the field of quantum
computing that motivated the development of the library.

\hypertarget{graph-creation-manipulation-and-traversal}{%
\subsection{Graph Creation, Manipulation, and
Traversal}\label{graph-creation-manipulation-and-traversal}}

The first use case is based on the manipulation of directed acyclic
graphs (DAGs) by Qiskit using \emph{rustworkx}. Qiskit represents
quantum circuits as DAGs on which the compiler operates on to perform
analysis and transformations (\protect\hyperlink{ref-Childs2019}{Childs
et al., 2019}).

\begin{figure}
\centering
\includegraphics[width=1\textwidth,height=\textheight]{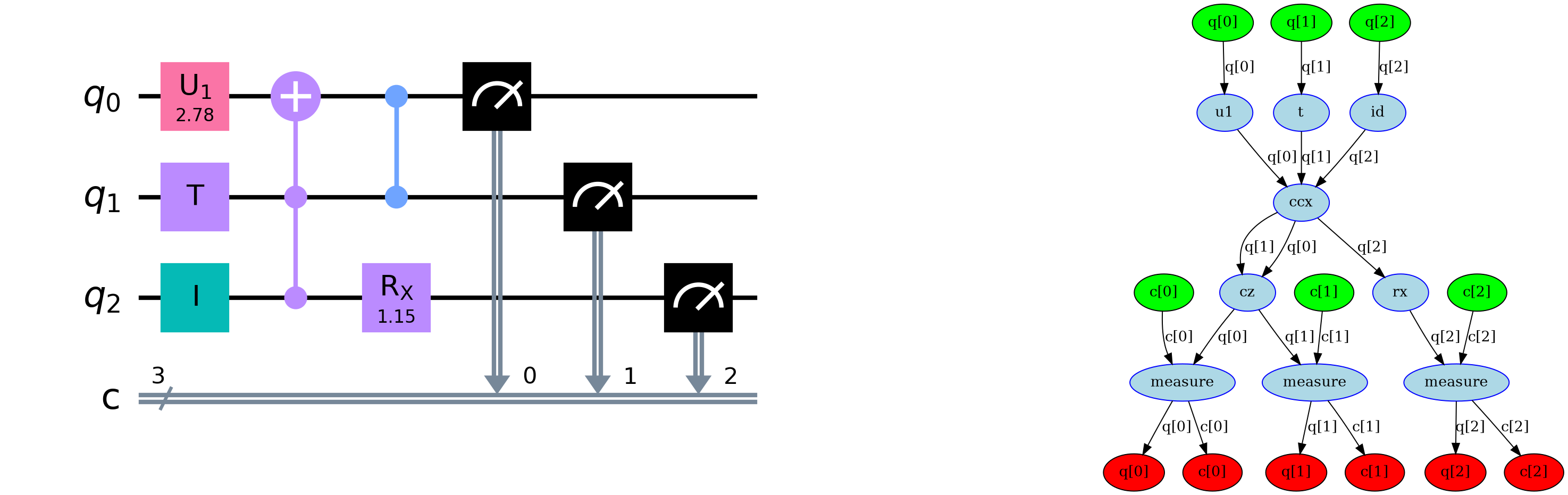}
\caption{Quantum circuit and its equivalent representation as a DAG of
instructions built by Qiskit.\label{fig:dagexample}}
\end{figure}

Qiskit creates a DAG with nodes that represent either instructions or
registers present in the quantum circuit
(\protect\hyperlink{ref-Cross2021}{Cross et al., 2021}) and with edges
that represent the registers each instruction operates on. Qiskit also
applies transformations to the instructions, which manipulate the graph
by adding and removing nodes and edges. \emph{rustworkx} brings the
graph data structure underlying those operations.

In addition, Qiskit needs to traverse the graph. Some transformations,
such as greedily merging instructions to reduce circuit depth, require
graph traversal. \emph{rustworkx} offers the methods for traversals such
as breadth-first search, depth-first search, and topological sorting.

\hypertarget{subgraph-isomorphism}{%
\subsection{Subgraph Isomorphism}\label{subgraph-isomorphism}}

The second use case is based on the qubit mapping problem for Noisy
Intermediate-Scale Quantum (NISQ) devices
(\protect\hyperlink{ref-Bharti2021}{Bharti et al., 2022};
\protect\hyperlink{ref-Preskill2018}{Preskill, 2018}). NISQ devices do
not have full connectivity among qubits, hence Qiskit needs to take into
account an undirected graph representing the connectivity of the device
when compiling quantum circuits. Qiskit transforms the quantum circuit
such that the pairs of qubits executing two-qubit gates respect the
device’s architectural constraints. There are many proposed solutions to
the qubit mapping problem, including algorithms based on subgraph
isomorphism (\protect\hyperlink{ref-Li2021}{Li et al., 2021}).

\begin{figure}
\centering
\includegraphics[width=0.525\textwidth,height=\textheight]{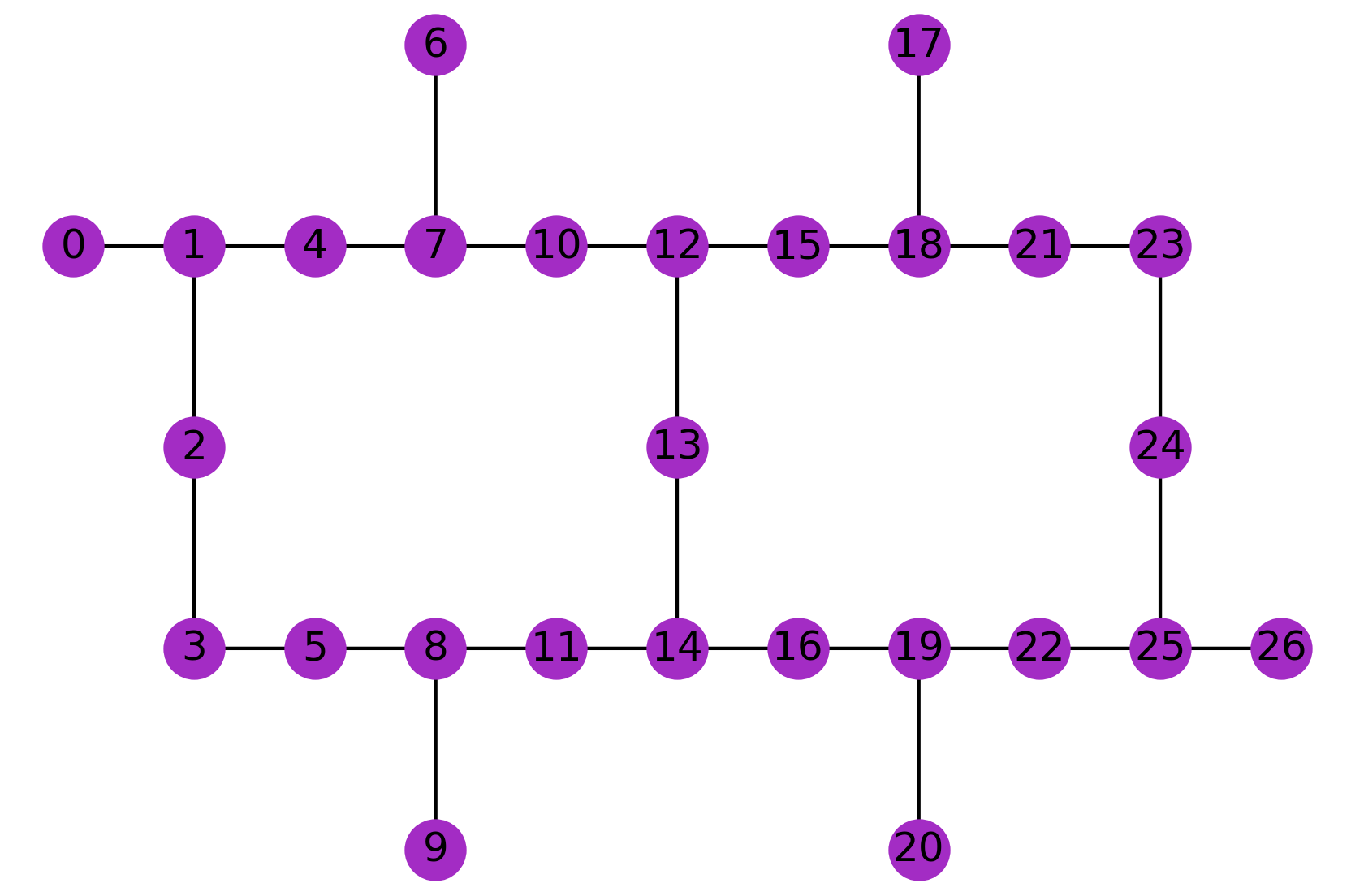}
\caption{Graph representing the connectivity of the
\textbf{\texttt{ibmq\_montreal}} device. Qiskit can check if the
required connectivity by a circuit is subgraph isomorphic to the
device’s connectivity when solving the qubit mapping
problem.\label{fig:graph_mtl_example}}
\end{figure}

\emph{rustworkx} implements the VF2 algorithm
(\protect\hyperlink{ref-Cordella2004}{Cordella et al., 2004}) and some
of the VF2++ heuristics (\protect\hyperlink{ref-Juttner2018}{Jüttner \&
Madarasi, 2018}) to solve subgraph isomorphism. The implementations
include both checking if a mapping exists and returning a mapping among
the nodes. Qiskit leverages \emph{rustworkx} to provide an experimental
layout method based on VF2. Qiskit checks if the graph representing the
connectvity required by the circuit and the connectivity provided by the
device are subgraph isomorphic. If they are, Qiskit uses VF2 mapping to
map the qubits without increasing the depth of the circuit.

\hypertarget{acknowledgements}{%
\section{Acknowledgements}\label{acknowledgements}}

We thank Kevin Krsulich for his help in getting \emph{rustworkx} ready
for use by Qiskit, Lauren Capelluto and Toshinari Itoko for their
continued support and help with code review, and all of the
\emph{rustworkx} contributors who have helped the library improve over
time.

\hypertarget{references}{%
\section*{References}\label{references}}
\addcontentsline{toc}{section}{References}

\hypertarget{refs}{}
\begin{CSLReferences}{1}{0}
\leavevmode\vadjust pre{\hypertarget{ref-Bergholm2020}{}}%
Bergholm, V., Izaac, J., Schuld, M., Gogolin, C., Alam, M. S., Ahmed,
S., Arrazola, J. M., Blank, C., Delgado, A., Jahangiri, S., McKiernan,
K., Meyer, J. J., Niu, Z., Száva, A., \& Killoran, N. (2020).
\emph{PennyLane: Automatic differentiation of hybrid quantum-classical
computations}. \url{https://doi.org/10.48550/ARXIV.1811.04968}

\leavevmode\vadjust pre{\hypertarget{ref-Bharti2021}{}}%
Bharti, K., Cervera-Lierta, A., Kyaw, T. H., Haug, T., Alperin-Lea, S.,
Anand, A., Degroote, M., Heimonen, H., Kottmann, J. S., Menke, T., Mok,
W.-K., Sim, S., Kwek, L.-C., \& Aspuru-Guzik, A. (2022). Noisy
intermediate-scale quantum algorithms. \emph{Reviews of Modern Physics},
\emph{94}(1). \url{https://doi.org/10.1103/revmodphys.94.015004}

\leavevmode\vadjust pre{\hypertarget{ref-bluss2021}{}}%
bluss, Comets, J.-M., Borgna, A., Larralde, M., Mitchener, B., \&
Kochkov, A. (2021). Petgraph. In \emph{GitHub repository}. GitHub.
\url{https://github.com/petgraph/petgraph}

\leavevmode\vadjust pre{\hypertarget{ref-Childs2019}{}}%
Childs, A. M., Schoute, E., \& Unsal, C. M. (2019). {Circuit
Transformations for Quantum Architectures}. In W. van Dam \& L.
Mancinska (Eds.), \emph{14th {C}onference on the {T}heory of {Q}uantum
{C}omputation, {C}ommunication and {C}ryptography (TQC 2019)} (Vol. 135, pp.
3:1–3:24). Schloss Dagstuhl–Leibniz-Zentrum fuer Informatik.
\url{https://doi.org/10.4230/LIPIcs.TQC.2019.3}

\leavevmode\vadjust pre{\hypertarget{ref-Cordella2004}{}}%
Cordella, L. P., Foggia, P., Sansone, C., \& Vento, M. (2004). A
(sub)graph isomorphism algorithm for matching large graphs. \emph{IEEE
Transactions on Pattern Analysis and Machine Intelligence},
\emph{26}(10), 1367–1372. \url{https://doi.org/10.1109/TPAMI.2004.75}

\leavevmode\vadjust pre{\hypertarget{ref-Cross2021}{}}%
Cross, A. W., Javadi-Abhari, A., Alexander, T., Beaudrap, N. de, Bishop,
L. S., Heidel, S., Ryan, C. A., Smolin, J., Gambetta, J. M., \& Johnson,
B. R. (2021). \emph{OpenQASM 3: A broader and deeper quantum assembly
language}. \url{https://doi.org/10.48550/arXiv.2104.14722}

\leavevmode\vadjust pre{\hypertarget{ref-Csardi2006}{}}%
Csardi, G., \& Nepusz, T. (2006). The igraph software package for
complex network research. \emph{InterJournal}, \emph{Complex Systems},
1695. \url{https://igraph.org}

\leavevmode\vadjust pre{\hypertarget{ref-SciPyProceedings_11}{}}%
Hagberg, A. A., Schult, D. A., \& Swart, P. J. (2008). Exploring network
structure, dynamics, and function using NetworkX. In G. Varoquaux, T.
Vaught, \& J. Millman (Eds.), \emph{Proceedings of the 7th {P}ython in
{S}cience {C}onference} (pp. 11–15).
\url{http://conference.scipy.org/proceedings/SciPy2008/paper_2/}

\leavevmode\vadjust pre{\hypertarget{ref-Hewitt2021}{}}%
Hewitt, D., Kanagawa, Y., Kim, N., Grunwald, D., Niederbühl, A.,
messense, Kolenbrander, B., Brandl, G., \& Ganssle, P. (2021). PyO3. In
\emph{GitHub repository}. GitHub. \url{https://github.com/PyO3/pyo3}

\leavevmode\vadjust pre{\hypertarget{ref-Jha2021}{}}%
Jha, S., Chen, J., Householder, A., \& Mi, A. (2021). Qtcodes. In
\emph{GitHub repository}. GitHub.
\url{https://github.com/yaleqc/qtcodes}

\leavevmode\vadjust pre{\hypertarget{ref-Juttner2018}{}}%
Jüttner, A., \& Madarasi, P. (2018). VF2++—an improved subgraph
isomorphism algorithm. \emph{Discrete Applied Mathematics}, \emph{242},
69–81. \url{https://doi.org/10.1016/j.dam.2018.02.018}

\leavevmode\vadjust pre{\hypertarget{ref-Leskovec2016}{}}%
Leskovec, J., \& Sosič, R. (2016). SNAP: A general-purpose network
analysis and graph-mining library. \emph{ACM Trans. Intell. Syst.
Technol.}, \emph{8}(1). \url{https://doi.org/10.1145/2898361}

\leavevmode\vadjust pre{\hypertarget{ref-Li2021}{}}%
Li, S., Zhou, X., \& Feng, Y. (2021). Qubit mapping based on subgraph
isomorphism and filtered depth-limited search. \emph{IEEE Transactions
on Computers}, \emph{70}(11), 1777–1788.
\url{https://doi.org/10.1109/tc.2020.3023247}

\leavevmode\vadjust pre{\hypertarget{ref-Matsakis2014}{}}%
Matsakis, N. D., \& Klock, F. S. (2014). The {R}ust language.
\emph{Proceedings of the 2014 ACM SIGAda Annual Conference on High
Integrity Language Technology}, 103–104.
\url{https://doi.org/10.1145/2663171.2663188}

\leavevmode\vadjust pre{\hypertarget{ref-Peixoto2014}{}}%
Peixoto, T. P. (2014). The graph-tool {P}ython library. \emph{Figshare}.
\url{https://doi.org/10.6084/m9.figshare.1164194}

\leavevmode\vadjust pre{\hypertarget{ref-Preskill2018}{}}%
Preskill, J. (2018). Quantum computing in the NISQ era and beyond.
\emph{Quantum}, \emph{2}, 79.
\url{https://doi.org/10.22331/q-2018-08-06-79}

\leavevmode\vadjust pre{\hypertarget{ref-Staudt2016}{}}%
Staudt, C. L., Sazonovs, A., \& Meyerhenke, H. (2016). NetworKit: A tool
suite for large-scale complex network analysis. \emph{Network Science},
\emph{4}(4), 508–530. \url{https://doi.org/10.1017/nws.2016.20}

\leavevmode\vadjust pre{\hypertarget{ref-Stone2021}{}}%
Stone, J., \& Matsakis, N. D. (2021). Rayon: A data parallelism library
for {R}ust. In \emph{GitHub repository}. GitHub.
\url{https://github.com/rayon-rs/rayon}

\leavevmode\vadjust pre{\hypertarget{ref-Sagemath2020}{}}%
The Sage Developers. (2020). \emph{{S}ageMath, the {S}age {M}athematics
{S}oftware {S}ystem ({V}ersion 9.1)}.
\url{https://doi.org/10.5281/zenodo.4066866}

\leavevmode\vadjust pre{\hypertarget{ref-Qiskit2021}{}}%
Treinish, M., Gambetta, J., Rodríguez, D. M., Marques, M., Bello, L.,
Wood, C. J., Gomez, J., Nation, P., Chen, R., Winston, E., Gacon, J.,
Cross, A., Krsulich, K., Sertage, I. F., Wood, S., Alexander, T.,
Capelluto, L., Puente González, S. de la, Rubio, J., … Woerner, S.
(2021). \emph{Qiskit/qiskit-terra: Qiskit terra 0.19.1} (Version 0.19.1)
{[}Computer software{]}. Zenodo.
\url{https://doi.org/10.5281/zenodo.2583252}

\leavevmode\vadjust pre{\hypertarget{ref-Ullberg2021}{}}%
Ullberg, S. (2021). {a}tompack: A flexible {P}ython library for atomic
structure generation. In \emph{GitHub repository}. GitHub.
\url{https://github.com/seatonullberg/atompack}

\end{CSLReferences}

\end{document}